\documentstyle [a4,epsfig]{paper}
\begin{document}

%


\vspace*{\stretch{1.0}}
 \begin{center}
    \Large\bf{Current voltage characteristics  and excess noise  at the trap filling transition in polyacenes.}\\
		\vspace*{\stretch{0.3}}
     \large{Jeremy Pousset$^{1}$, Eleonora Alfinito$^{2,3}$, \\ 
		Anna Carbone$^{4}$, Cecilia Pennetta$^{1}$, Lino Reggiani$^{1}$} \\
		\vspace*{\stretch{0.1}}
  \small\it{$^{1}$Dipartimento di Matematica e
Fisica, "Ennio de Giorgi" Universit\`a del Salento, via Monteroni
73100 Lecce, Italy, 
\\$^{2}$Dipartimento di Ingegneria dell' Innovazione \\ Universit\`a del
Salento, via Monteroni, I-73100 Lecce, Italy \\
$^{3}$INFN (Istituto Nazionale di Fisica Nucleare), Sezione di Lecce - Lecce, Italy\\
$^{4}$Physics Department,  Politecnico di Torino, Corso Duca degli Abruzzi 24, 10129 Torino, Italy}
	\end{center}
  \vspace*{\stretch{0.2}}

\begin{abstract}
Experiments in organic semiconductors (polyacenes)  evidence a
strong super quadratic increase of the current-voltage (I-V) characteristic at voltages in the transition region between linear (Ohmic) and  quadratic (trap free space-charge-limited-current) behaviours.
Similarly, excess noise measurements at a given frequency and increasing voltages evidence  a sharp peak of the relative spectral density of the current noise in concomitance with the strong super-quadratic I-V characteristics.
Here we discuss the physical interpretation of these experiments
in terms of an essential contribution from  field assisted trapping-detrapping processes of injected carriers.
To this purpose, the fraction of filled traps determined by the I-V characteristics is used to evaluate the excess noise in the trap filled transition (TFT) regime.
We have found an excellent agreement between the predictions of our model and existing experimental results in tetracene and pentacene thin films of different length in the range $0.65 \div 35 \ \mu m$.
\end{abstract}
%
%
\section{Introduction}
Organic devices, based on polymeric materials or molecular semiconductors, succesfully compete with traditional electronic devices at least in terms of cost, flexibility and weight \cite{muccini,fleissner,chen,lezzi15,song}.
The performance of organic devices is controlled by charge carriers that are injected at the molecule-metal interfaces.
In turns, injected carriers are drastically affected by the presence of
trapping centers related to defect states.
The effect of thermal and electrical stresses on charge carrier trapping and detrapping (TD) processes is widely investigated in the literature
\cite{lang,miya,koch,yang,knipp,dinelli,schwalb,boer,kang,chandra,giulia}, together with studies devoted to noise in organic semiconductors
\cite{carbone05,jurchescu08,carbone09,lezzi15,song,song1}.
\par
In particular, transport  measurements in polyacenes  have evidenced a strong superlinear increase of the
current-voltage (I-V) characteristics \cite{boer,carbone05,lezzi15}, and  associated noise measurements \cite{carbone05,lezzi15} have shown a sharp peaking of the relative spectral-density of excess current-noise in concomitance with the superlinear increase of the I-V characteristics.
This noise peaking occurs at voltage regions corresponding to the crossover between Ohmic and space-charge-limited-current
(SCLC) regimes \cite{carbone05}, at the so called trap-filling transition (TFT).
The interpretation of the experiments for the case of tetracene was previously addressed in terms of trapping-detrapping processes of the injected carriers by a single level of deep traps 
\cite{carbone09}.
Here we generalize the single trap model to the case of the presence of several trap levels, as it often occurs  under SCLC conditions, and of variable sample lengths. 
To validate the model, we consider existing  measurements performed on tetracene and  pentacene films of length in the range $0.65 \div 30 \ \mu m$ 
\cite{boer,carbone05}.
The agreement between theory and experiments provides a series of physical parameters that characterize the traps present in different materials. 
\par
The paper is organized as follows.
The theoretical model which, following \cite{carbone09}, is extended to the presence of  many levels of traps, is developed in the next Sec. 2.
Section 3 reports the comparison of theoretical calculations with existing experiments carried out in tetracene and pentacene samples of diferent lengths.
Here, the mechanisms of voltage enhanced detrapping, as quantified by the fraction of ionized  traps, are identified and discussed. 
Major conclusions are drawn in Sec. 4.
\section{Theoretical model }
We consider I-V characteristics and excess noise of a two-terminal sample with length $L$ and cross-sectional area $A$ characterized by the presence of a fully ionized shallow trap level and a set of $N$ independent traps levels each with a given energy,  thus generalizing previous results  carried out for a single-trap level \cite{carbone09}.
Accordingly, within a phenomenological approach,  we develop the appropriate expressions for the transport and  relative excess current-noise spectrum as function of an applied voltage.
We notice, that according to the phenomenological model all macroscopic physical quantities are considered as spatially averaged ones, thus microscopic spatial dependence of fields, traps and free carriers are implicitly accounted for by the voltage dependence of the considered quantities.  
\subsection{Transport}
Following \cite {lampert70}, charge transport is assumed to consist of  three kinds of  regimes, namely: Ohmic (linear I-V at the lowest  applied voltages), TFT through a sequence of $N$ trap levels, each level being  responsible for a sharp superquadratic I-V chracterics at intermediate applied voltages, and an SCLC quadratic I-V at further increasing applied voltages characterized  by a trapping factor $\Theta_i \le 1$. The asymptotic value $\Theta_i = 1$
corresponds to the condition of trap free SCLC \cite{lampert70}.
\par
The fit  of the experimental I-V characteristics is obtained as the sum of the currents respectively in:
(i)  the Ohmic regime; (ii) the TFT regime pertaining to a set of $1 \le i \le N$trap levels, which are weighted by the fraction of filled traps $u_i (V)$  that are functions of the applied voltage and are limited within the values $0 \le u_i(V) \le 1$; (iii) a quadratic SCLC regime that is included in the expression of the TFT current when $u_i=1$.
Here traps are assumed to be uniformly distributed over space while  the inhomogeneous spatial distribution of the electric field and charge carriers are accounted for by the nonlinear behaviour of the I-V characteristics at increasing values of the applied voltage.
According to this decomposition, we obtain:
\begin{equation}
I=I_{\Omega} + I_{TFT} 
\end{equation}
with
\begin{equation}
I_{\Omega} = \frac{V}{R_{\Omega}}
\end{equation}
where 
$$
R_{\Omega} = \frac{L}{A e n_0 \mu}
$$ 
is the Ohmic resistance, $e$ is the unit charge, $n_0$ is the free carrier thermal concentration, and $\mu$ is the carrier mobility that is assumed to be independent of the applied voltage.
\begin{equation}
 I_{TFT} = \sum_{i=1}^N I_{TFT_i} = \sum_{i=1}^N u_i I_{SCLC_i}
\end{equation}
where $I_{SCLC_i}$  is obtained from the Mott-Gurney law \cite{lampert70} in the presence of a trap level, $i$, with concentration  $n^i_{t,0}$ of trapping centers and concentration $n_i$ of free carriers coming from the sum of thermally activated $n_0$ and injected $n_{in}$
as
\begin{equation}
I_{SCLC_i}= \frac{9A \epsilon_0 \epsilon_r \mu \Theta_i  V^2}{8 L^3}
\end{equation}
where $ \epsilon_0$ is the vacuum permittivity, $ \epsilon_r$ is the relative dielecric constant of the material, $\Theta_i = n_i / n^i_{t,0}$ the trapping factor.
Notice that the  carrier concentration injected from the contact $n_{in}$ is given by:
\begin{equation}
n_{in} = \frac{\epsilon_0 \epsilon_r V}{e L^2}\Theta_i
\end{equation}
For convenience we can also write
\begin{equation}
I_{SCLC_i}=I_{\Omega} \frac{9 n_{in}}{8n_0} 
\end{equation}   
We notice that, when $\Theta_i=1$,  $I_{SCLC}$ is the current corresponding to the quadratic trap-free SCLC regime, the maximum current that the sample can support.
\par
For the fitting with experiments,
the value of $I_{\Omega}$ is obtained from the Ohmic regime reported in Eq. (2)
and the value of $I_{SCLC_i}$ is taken from the $i$-th SCLC regime extrapolated at the corresponding flexing points of the I-V characteristics.
Accordingly, for the case of the first trap the $u_{1}=u_{1}(V)$ is obtained by best fitting the full experimental curve $I_{exp}=I_{exp}(V)$ using
\begin{equation}
u_{1}(V) = \frac{I_{exp}-I_{\Omega}}{I_{SCLC_1}}
\end{equation}
in the range of voltages for which $0 < u_1 < 1$.
\par
For the case of a second trap the $u_{2}=u_{2}(V)$ is obtained by best fitting the full experimental curve $I_{exp}=I_{exp}(V)$ as
\begin{equation}
u_{2}(V) = \frac{I_{exp}-I_{\Omega}-I_{SCLC_1}}{I_{SCLC_2}}
\end{equation}
in the range of voltages for which $0 < u_2 < 1$
and so on for all the other possible traps.
\par
The value of $I_{SCLC}$  correponding to the trap free SCLC regime is taken by extrapolating the current value at $V \rightarrow \infty$.
\par
The theoretical value of the voltage dependence of the fraction of ionized traps, $ u_i^{th}$ is taken by using two alternative models.
The first is the  Quasi-Fermi (QF) model of the form:
\begin{equation}
u_i^{th,QF} = \frac{1}{1+ g \exp[(\Delta \epsilon_{QF} - \gamma_{QF} e V)/(k_BT)]}
\end{equation}
where $g=4$ is the degeneracy of the trap level for the case of acceptors, $\Delta \epsilon_{QF} =(\epsilon_F -\epsilon_i)_{QF}$, with $\epsilon_F $ the Fermi level at thermal equilibrium and $\epsilon_i$ the $i$-th trap energy level,
$\gamma_{QF} = 1/(L n_t^{1/3})$ is a numerical fitting parameter relating the mean value of  trap concentration  $n_t$ with the sample length $L$,  $k_B$ is the Boltzmann constant and $T$ is the bath temperature.
\par
The second model considers the voltage dependence of the fraction of filled traps governed by the Poole-Frenkel (PF) effect \cite{hartke68} in the form:
\begin{equation}
u_i^{th,PF} = \frac{1}{1+ g \exp[(\Delta \epsilon_{PF} -  e \beta_{PF}( V /L)^{1/2}) /(k_BT)]}
\end{equation}
with $\beta_{PF}$
an adjustable Poole-Frenkel factor fitted to experiments and found to be higher for about one order of magnitude than the standard value $[e / (4 \pi \epsilon_r \epsilon_0)]^{1/2} \approx 2.0 \times 10^{-5} \ (Vm)^{1/2}$.
\par
The values of the low field mobility $\mu$, thermal carrier density $n_{0}$, $\Delta \epsilon_{PF,QF}$ and $\Theta_i$ should be taken consistently with critical values of the I-V  characteristics as will be detailed in Sec. 3 where a comparison between theory and experments will be carried out.
\subsection{Noise}
According to the current decomposition in Eqs. (1) to (4), the relative excess
current-noise is written as:
\begin{equation}
S(f,V) =  S_{\Omega}(f) +  S_{TFT}(f,V) +  S_{SCLC}(f,V) \label{eq:nois_tot}
\end{equation}
where, according to Hooge formula \cite{hooge69}, the Ohmic component of
the $1/f$ noise is:
\begin{equation}
S_{\Omega}(f) = \frac{\alpha_{\Omega}}{A L n_0 f}
\end{equation}
with
$\alpha_{\Omega}$ the Ohmic Hooge parameter \cite{hooge69} (as well known, the Ohmic component is independent of applied voltage).
\par
For  TFT  noise from N independent trap levels we take a superposition  of simple 
Lorentzians as \cite{vanderziel70}
\begin{equation}
S_{TFT}(f,V) = \sum_{i=1}^N S_{TFTi} =\sum_{i=1}^N B_i(f)  u_i(V)[1-u_i(V)]
\end{equation}
with
\begin{equation}
B_i(f)= \frac{4}{N_{ti}} \frac{\tau_i }{1 + (2 \pi f \tau_i)^2}
\end{equation}
the amplitude of each $i$-th trapping-detrapping source assumed to be independent from each other,  where $N_{ti}$ is the total number of $i$-th traps in the volume of the device, and $\tau_i$ is the associated carrier  life-time.
We notice that at the onset of the TFT regime the number of free carriers  equals the number of filled traps. 
\par
The Lorentzian spectrum is then generalized to the case that the single lifetime is broadened in a sufficient wide range of values so that the Lorentzian can orginate a $1/f$ spectrum in the correspoding frequency region as:
\begin{equation}
B_i^{br}(f)= P_i \frac{\arctan(2 \pi f \tau_{2,i}) - \arctan(2 \pi f \tau_{1,i} )}{2 \pi f}
\end{equation}
with $P_i$ a constant parameter to be fitted by comparison with experiments.
\par
Accordingly, when the two time constants  $\tau_{2,i}$ and $\tau_{1,i}$ defining the broadenong region $\tau_{1,i} \le \tau_i \le \tau_{2,i}$,
satisfy the condition $2 \pi f \tau_{2,i} \gg 1$ and  $2 \pi f \tau_{1,i} \ll 1$,  respectively, the two $\arctan$ terms in the numerator of Eq. (15) are approximately equal to $π/2$ and $0$, and inside the broadenng region the superposition of exponential relaxation processes  gives rise to a $1/f$ spectrum:
\begin{equation}
B_i^{br}(f)=\frac{P_i}{4 f}
\end{equation}
\par
By generalizing the trap-free SCLC  Kleinpenning formula \cite{klein} to the case in which traps are present, the SCLC$_i$ component of the $1/f$ noise is taken as:
\begin{equation}
S_{SCLC_i}(f,V) =
\sum_{i=1}^N  \frac{4eL \alpha_{SCLC_i} u_i^2} {5A \epsilon_0 \epsilon_r \Theta_i f V }
\end{equation}
with $\alpha_{SCLC_i}$ an SCLC$_i$  Hooge parameter.
\par
Accordingly, the $1/f$ noise sources in Eqs. (12) and (16) are attributed to mobility fluctuations.
\section{Comparison with experiments}
The theoretical model developed above is applied to experimental data obtained on tetracene and pentacene samples \cite{boer,carbone05}.
In so doing, we further validate and generalize the procedure presented in Ref. \cite{carbone09}.
The values of the low field mobility $\mu$, thermal carrier density $n_{0}$ and $\Theta_i$ should be taken consistently with the asymptotics values of the $I(V)$ characteristics.
The simultaneous fitting of the I-V charateristics and of the relative excess-noise component at the given frequency will help in providing estimates of the trap concentration and the carrier lifetime in a consistent way.
\subsection {Current-voltage characteristic}
\begin{table}[pt]
\caption{Relevant parameters for tetracene samples \cite{boer}.
The lower (higher) values of mobility refers to the shorter (longer) sample length.} 
\begin{tabular}{ccc} \hline
Relative dielectric constat & $\epsilon_r$ & $3$ \\ 
Zero-field hole mobility  & $\mu$ & $(0.014 - 0.59 ) \ \mathrm{cm^{2}/ (s V)}$ \\
Trapping factor & $\Theta$ & $1.3 \times 10^{-6} \div 1 $ \\
Thermal free carrier concentration &$n_0$ & $\approx 0 - 6.3 \times 10^{9}  \mathrm{ cm^{-3}}$ \\
Density of traps &	$n_{t,0}$ & $1.0 \times 10^{15} \ \mathrm{ cm^{-3}}$ \\
----& $\Delta \epsilon_{QF,PF}$ & $\approx 0.51 \div 1.01 \ \mathrm{eV}$\\
\hline
\end{tabular}
\end{table}
Figures 1, 2 and Figures 3, 4 report, respectively, the I-V characteristics and the associated fraction of filled traps for the case of two different rectangular  samples of tetracene \cite{boer}.
Data reported in Figs. 1 and 2 refer to a sample width of cross sectional area $A = 0.028 \ mm^2$ and length $L = 0.03 \ mm$, respectively.
In this case, the fitting between theory and experiments is obtained by taking:

\begin{equation}
I_{\Omega}=0
\end{equation}

\begin{equation}
I_{SCLC1}=1.3 \times 10^{-3} \ V^2 \ pA
\end{equation}

\begin{equation}
I_{TFT2}=5.0 \times 10^3 \ V^2 \ u(V) \ pA
\end{equation}
which imply $\Theta_1=1,3 \times 10^{-6}$, $\Theta_2=1$, 
$\mu=0.014 \ cm^2/(sV)$
\par
Data reported in Figs. 3 and 4  refer to a sample of cross sectional area $A = 0.028 \ mm^2$ and length $L = 0.025 \ mm$, respectively.
In this case, the fitting betwen theory and experiments is obtained by taking:
\begin{equation}
I_{\Omega}=1.5 \times 10^{-2} \ V \ pA
\end{equation}
\begin{equation}
I_{TFT1}=1.2 \times 10^{-2} \ V^2 \  u_1(V)  \ pA
\end{equation}
which implies $\Theta_1=1$, $n_0 = 1.9 \times 10^7 cm^{-3}$, 
$\mu=0.59 \ cm^2/(sV)$
\par
Dashed curves in Fig. 1 and  Fig. 3 refer to the theoretical fitting carried out within the model of Sec. 2 for a single trap level  with the fraction of filled traps reported in Figs. 2 and 4 for the PF model.
In Figs. 2 and 4  symbols refer to the values extracted from the fit of experiments and curves refer to  the theoretical results obtained from statistics using a Quasi-Fermi (QF) model  or a Poole-Frenkel (PF) model, respectively.
We found that QF and PF models give very similar results, with  the PF model
providing a sligthly better agreement with experiments.
\par
The parameters extracted from the fitting for the tetracene samples are summarized in Table 1.
\par
Figures 5 and 6  report, respectively, the I-V characteritics and the associated fraction of filled traps obtained on a sample of pentacene \cite{carbone05} with $A = 0.1 \ cm^2$ and $L = 0.85 \ \mu m$.
In this case, the fitting is obtained by taking:
\begin{equation}
I_{\Omega}=13 \ V \ pA
\end{equation}
\begin{equation}
I_{TFT1}=7.0 \times 10^3 \  V^2 \ u_1(V)  \ pA
\end{equation}
\begin{equation}
I_{TFT2}=2.0 \times 10^5 \  V^2 \ u_2(V)  \ pA
\end{equation}
which implies $\Theta_1=3.5 \times 10^{-2}$, $\Theta_2=1$, 
$\mu=4.1 \times 10^{-6} \ cm^2/(sV)$
\par 
The continuous curve in Fig. 5  refers to the theoretical fitting carried out within the model of Sec. 2 for a double trap level with the fraction of filled traps reported in Fig. 6.
In Figure 6 symbols refer to the values extracted from the fit of experiments and curves refer to  the theoretical results obtained from statistics using a QF model or a PF model, respectively.
Even in this case, we found that QF and PF models give very similar results, with  the PF model providing a sligthly better agreement with experiments.
\subsection {Relative excess current-noise}
Here we consider the relative excess-noise characteristics measured in tetracene and pentacene samples of \cite{carbone05}.
For the case of tetracene, Ref. \cite{carbone09}  already reported the fit of the I-V charateristics and of the relative excess-noise measured at 20 Hz as function of the applied voltage for a Au/Tc/Al sample of length $0.85 \ \mu m$ and cross-sectional area of $0.1 \ cm^2$.
Below, noise spectra  in the measured range $1 \div 10^4 \ Hz$ of the same tetracene sample are fitted using the frequency expressions at the given voltage given by
\begin{equation}
S_{\Omega}(f)= \frac{36}{f} \  ps
\end{equation}
\begin{equation}
S_{TFT}(V,f) = 8.2 \times 10^2 \ \frac{ u(V)[1-u(V))] }{1 + (2 \pi f \tau)^2} \ ps
\end{equation}
for a single lifetime (Lorenzian) model, or
\begin{equation}
S_{TFT}(V,f) = 8.2 \times 10^3 \ \frac{ u(V)[1-u(V))] }{f} \ ps
\end{equation}
for a supeerposition of relaxation processes describewd by broadened set of lifetimes in the range $10^{-6} \div 10^2 \ s$, so that the broadening factor is well approximated by 1 in the considered frequency region.
\begin{equation}
S_{SCLC}(V,f) = \frac{800 \ u^2}{V f } \ ps
\end{equation}
Figure 7 reports the fraction of filled traps deermnwd from the fitting of the I-V experiments and that are here used for the fitting of the noise spectra.
\par
Figures 8 to 12 reports the noise spectra of tetracene \cite{carbone05}
for several voltages covering the range from  0.5 to 6  V.
We recall that the spectrum at 0.5 V corresponds to the Ohmic regime, spectra at 0.8 and 1.5 V corespond to the TFL regime, and spectra at 4.3 and 6 V correspond to the trap-free SCLC regime.
In particular, Fig. 9 at 0.8 V compares the shape of the spectra when going from a single lifetime, $\tau=8 \ ms$, to the broadened set of lifetimes responsible for the 1/f slope.
The case of broadened lifetimes well reproduce the $1/f$ like spectra exhibited by experiments in the full frequency range $1 \div 10^4 \ Hz$ and at voltage regions where the TFT regime is dominant.
We notice, that at the highest voltages experiments exhibit a spectra with a slope sligthy sharper than  the simple 1/f, and which is responsible of most of the misfit with theory. 
\par
For the case of pentacene \cite{carbone05}, the theoretical spectra of all the noise source considered are taken to exhibit an 1/f shape and the fitting with experiments is carried out at the frequency of 20 Hz by taking:
\begin{equation}
S_{\Omega}= \frac{30}{f}  \  ps
\end{equation}
\begin{equation}
S_{TFT1}(V,f) = \frac{3.6 \times 10^3  \ u1(V)[1-u1(V))]}{f}  \ ps
\end{equation}
\begin{equation}
S_{TFT2}(V,f) = \frac{4.8 \times 10^4 \ u2(V)[1-u2(V)]}{f} \  ps
\end{equation}
\begin{equation}
S_{SCLC1}(V,f) = \frac{3.8 \times 10^3 \ u_1^2}{V f } \ ps
\end{equation}
\begin{equation}
S_{SCLC2}(V,f) = \frac{3.8 \times 10^3 \ u_2^2}{V f}  \ ps
\end{equation}
The values of the fraction of filled traps obtained from the I-V fit are used as input parameters for the corresponding fit of the relative excess-noise at 20 Hz as function of the applied voltage whivh is reported in Fig. 13.
To fix the main parameters of the fitting between theory and experiments we have paralleled the procedure used in  \cite{carbone09}.
In particular, the maximum value of the trap concentrations and the corresponding estimate of a carrier free-time is obained by solving Eq. (14) for real values of $\tau$.
\par
The parameters extracted from the fitting of pentacene data are summarized in Table 2.
\begin{table}[pt]
\begin{center}
\caption{Relevant parameters for pentacene.
The given range of values are in correspondence with that of a contact surfacce of, respectively $10^{-1}$ or $10^{-5} \ cm^2$.}
\begin{tabular}{cccc} 
\hline
Relative dielectric constant & $\epsilon_r$ & $4.3$ \\ 
Zero-field hole mobility  & $\mu$ & $2.9 \times (10^{-6} \div  10^{-2})  \ \mathrm{cm^{2}/ (s V)}$ \\
Thermal free-carrier concentration &$n_0$ & $2.4 \times (10^{10} \div  10^{6})\  \mathrm{ cm^{-3}}$ \\
Density of traps &	$n^1_{t,0}$ & $3.2 \times (10^{10} \div 
 10^{14}) \ \mathrm{ cm^{-3}}$ \\
Density of traps &	$n^2_{t,0}$ & $8.8 \times (10^{10} \div  10^{14}) \  \mathrm{ cm^{-3}}$ \\
Valence band state density 	& $n_{v}$ & $10^{21} \ \mathrm{ cm^{-3}}$ \\
Carriers 1 free time &$\tau_1$ & $10^{-6} \div 10^2 \ s$ \\
Number of carriers 1 &$N_{t1}$ & $< 3.5 \times 10^7$ \\
Carriers 2 free time &$\tau_2$ & $10^{-6} \div 10^2 \ s$ \\
Number of carriers 2 &$N_{t2}$ & $< 6.6 \times 10^9$ \\
Deep trap energy level  & $\epsilon_{1,2}$ & $\approx 0.85 \div 1 \ \mathrm{eV}$ \\
-----& $\Delta \epsilon_{QF,PF}$ & $\approx (0.28 \div 0.52) \ \mathrm{eV}$ \\
Trapping factors & $\Theta_{1,2}$ & $3.5 \times (10^{-2}, \ 1) $ \\
Ohmic Hooge parameter & $\alpha_{\Omega}$ & $6.1 \times 10^{-6}$ \\
SCLC1 Hooge parameter & $\alpha_{\mathrm{SCLC_1}}$ & $2.2 \times ( 10^{-1} \div 10^{-5})$\\ 
SCLC2 Hooge parameter & $\alpha_{\mathrm{SCLC_2}}$ & $6.2 \times ( 1 \div 
10^{-4})$\\  
\hline
\end{tabular}
\end{center}
\end{table}
Here, in view of the extremely low value of the hole mobility that is estimated from the given geometry, we have considered the possibility that from the electrical point of view the effective area of the contacts be a factor of $10^{-4}$ smaller than the geometrical value reported in experiments \cite{carbone05}.
This can happen due to the strong inhomegenety of the organic material that can make only a small fraction of the contact area permeable to the current flow.
Accordingly, the corresponding parameters span a comparable range of values (see Table 2).
\section{Conclusions}
We have developed a phenomenological model that provides a quantitative interpretation of the current-voltage characteristic and the relative excess current-noise in the presence of space-charge limited conditins due to the presence of multiple trapping centers. 
The model is applied to the case of polyacenes where different sets of experiments are available from literature.
We have found an excellent agreement between the predictions of our model and experimental results in tetracene and pentacene thin films of different lemgth in the range $0.65 \div 35 \ \mu m$.
The agreement allows us to state that the sharp peak of noise in the TFT region exhibited by pentacene films arises from the
fluctuating occupancy of the traps due to trapping-detrapping processes.
The fitting of the I-V and the noise experiments extends  over 10 and 4 orders of magnitude, respectively, and  provides a set of  parameters (see Tables 1 and 2) of valuable interest for the characterization of the samples under investigation.
\par
Finally, we remark that the measured current noise spectrum in the voltage region controlled by the TFT was usually found to be $1/f$-like \cite{necludiov00,martin00,vandamme02,carbone05,ke08,kang11,xu11,harsh15,ciofi17}, thus without a direct evidence of a Lorentzian spectrum, as assumed in Eq. (14).
As a consequence, a broadening of the trap lifetimes in the range $10^{-6} < \tau < 10^2 \ s$ is considered to account for the $1/f$ shape of the noise spectra \cite{klein,penne,fleetwood15}.
%
%

%

\begin{figure}
 \begin{center}
  \includegraphics[width=9cm]{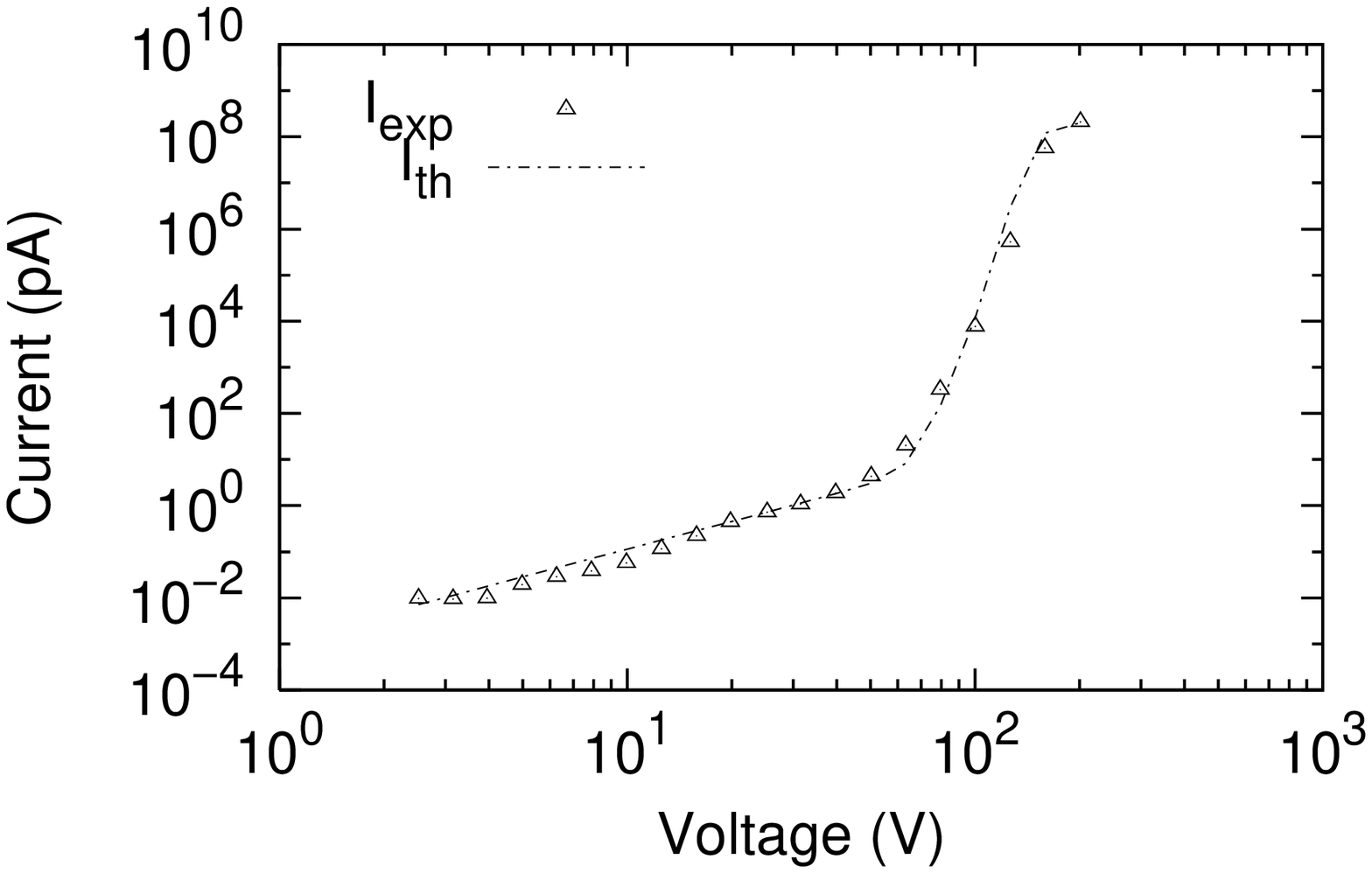}
 \end{center}
 \caption{ \label{fig1}
Current voltage characteristics of the tetracene  sample of length $25 \ \mu m$ at $T=300 \ K$ \cite{boer}.
Symbols refer to experiments and the line to the best fit obtained using the fraction of filled traps calculted within the PF model as reported in Fig. 2.
}
\end{figure}
\begin{figure}
 \begin{center}
  \includegraphics[width=9cm]{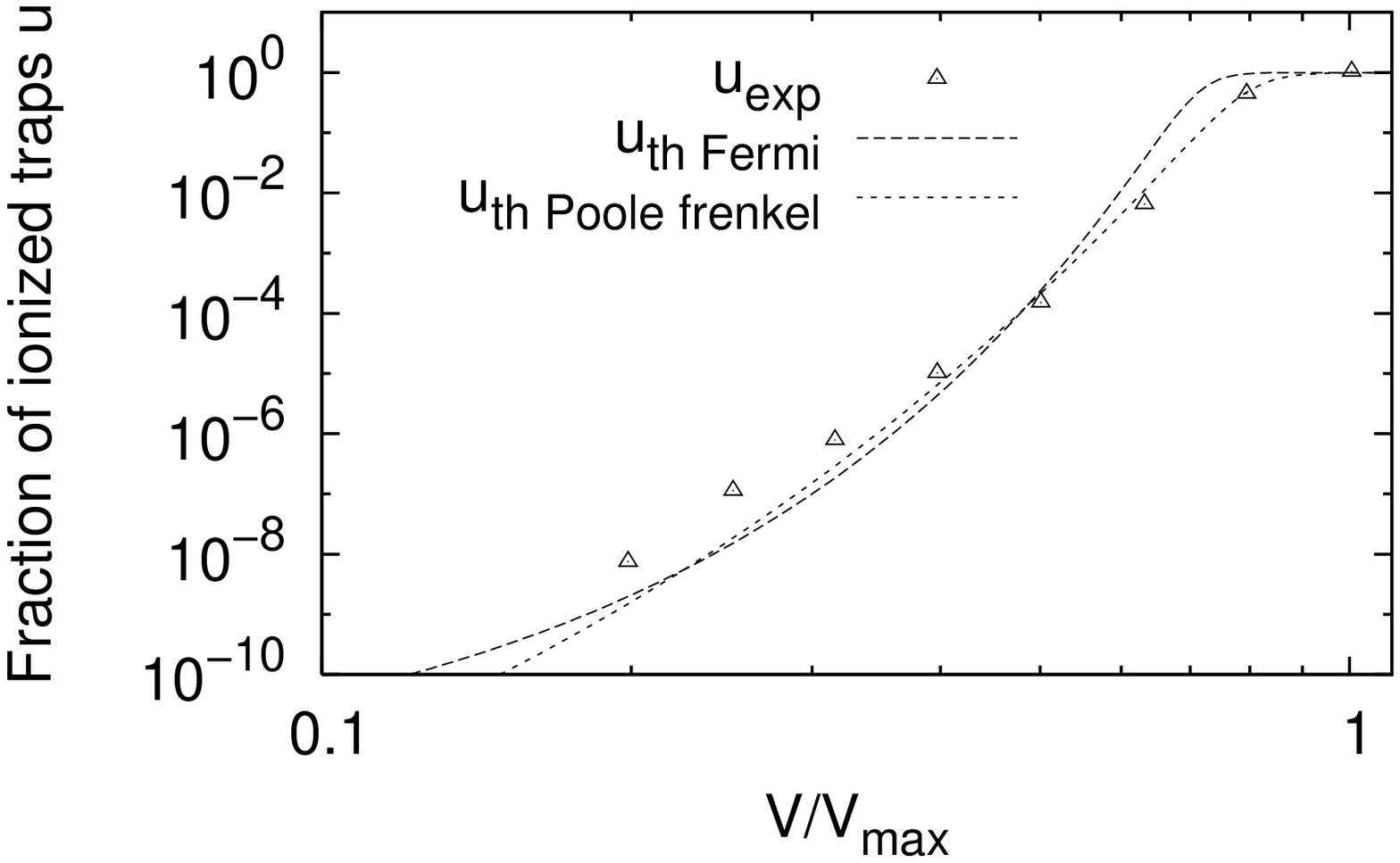}
 \end{center}
 \caption{ \label{fig2}
Fraction of filled traps of the sample in Fig. 1 as function of the applied voltage normalized to the value $V_{max} = 200 \ V$ corresponding to the voltage when $u=1$.
Symbols refer to the experimental vaòues deduced using Eq. (8)  from the I-V data in Fig. 1  and lines to the fit using the statistics with the QF and PF models, resperively.
Here $\gamma_{QF}=6.0 \times 10^{-3}$, $\Delta \epsilon_{QF} =0.67 \ eV$, and  $\Delta \epsilon_{PF} =1.01 \ eV$, $\beta_{PF}=4.2 \times 10^{-4} \ (Vm)^{1/2}$.
}
\end{figure}
\begin{figure}
 \begin{center}
  \includegraphics[width=9cm]{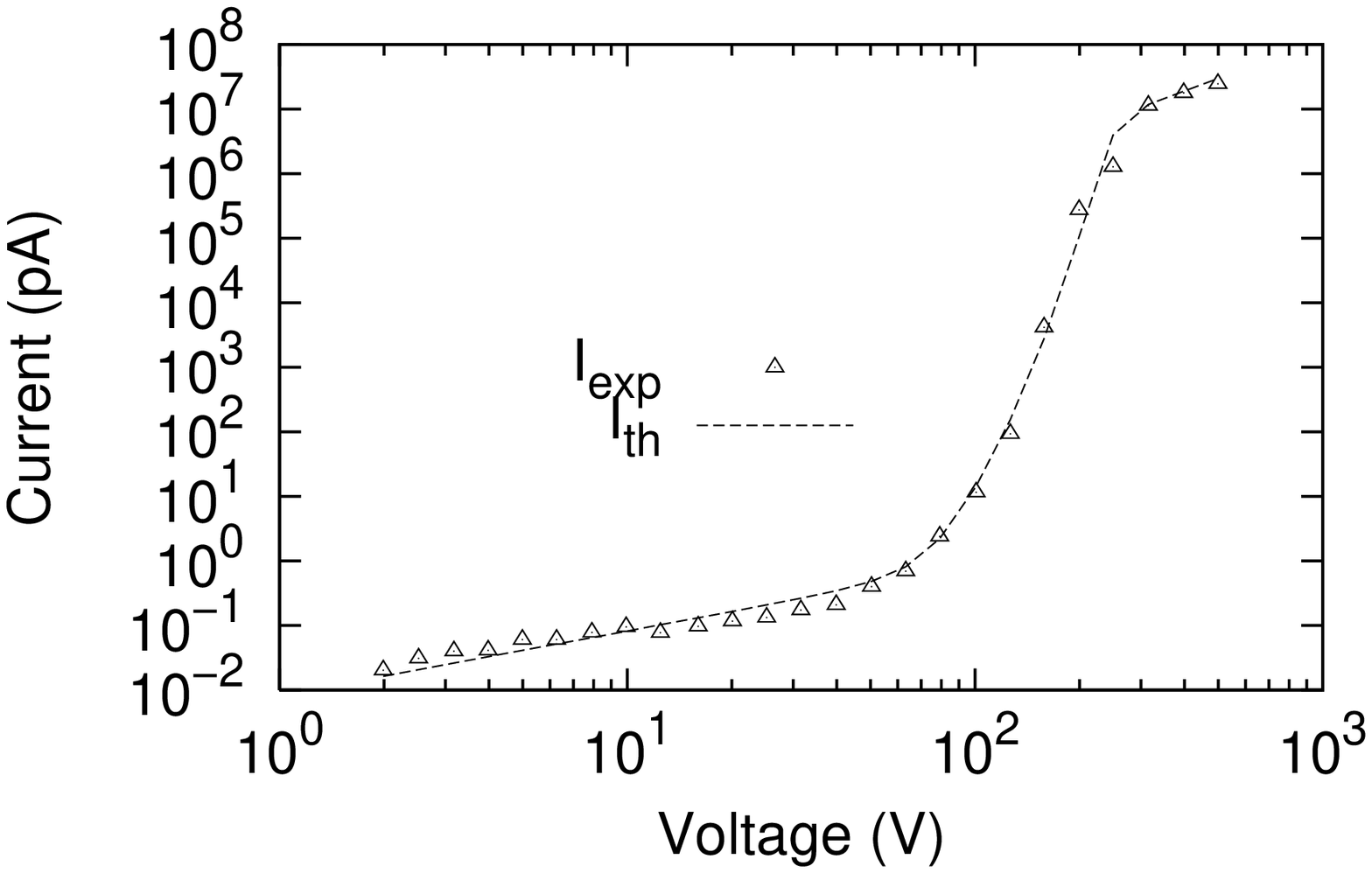}
 \end{center}
 \caption{ \label{fig3}
Current voltage characteristics of a tetracene  sample of length $30 \ \mu m$ $T=300 \ K$  \cite{boer}.
Symbols refer to experiments and the line to the best fit obtained using the fraction of filled traps calculted within the PF model as reported in Fig. 4.
}
\end{figure}
\begin{figure}
 \begin{center}
  \includegraphics[width=9cm]{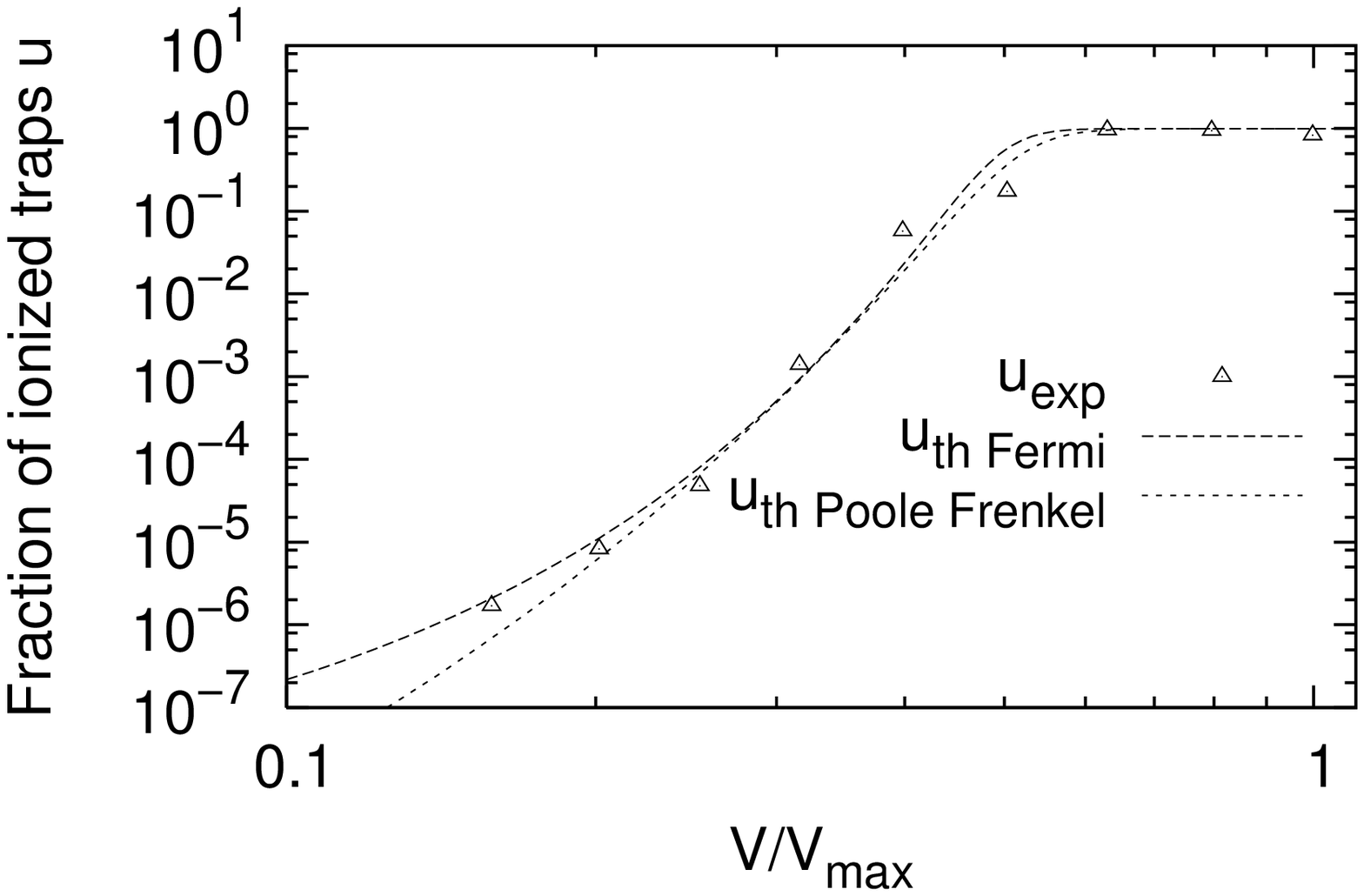}
 \end{center}
 \caption{ \label{fig4}
Fraction of filled traps of the sample in Fig. 3 as function of the applied voltage normalized to the value $V_{max} = 500 \ V$ corresponding to the voltage when $u=1$.
Symbols refer to the experimental values deduced using Eq. (8)  from the I-V data in Fig. 3  and lines to the fit using the statistics with  QF and PF models, respectively.
Here $\Delta \epsilon_{QF} =0.51 \ eV$, $\gamma_{QF}= 2.3 \times 10^{-3}$, and  $\Delta \epsilon_{PF} =0.78 \ eV$, $\beta_{PF}=2.8 \times 10^{-4} \ (Vm)^{1/2}$.
}
\end{figure}
\begin{figure}
 \begin{center}
  \includegraphics[width=9cm]{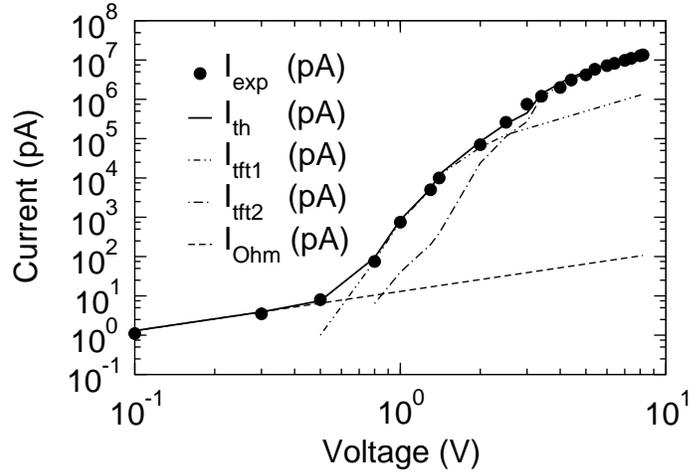}
 \end{center}
 \caption{ \label{fig5}
Current voltage characteristics of a pentacene  sample of length $0.85 \ \mu m$  at $T=300 \ K$ \cite{carbone05}.
Symbols refer to experiments, continuous line to the best fit obtained using the fraction of filled traps calculted within the Poole Frenkel model as reported in Fig. 6, dashed lines to the three components in which the total current is decomposed, see text.
}
\end{figure}
\begin{figure}
 \begin{center}
  \includegraphics[width=9cm]{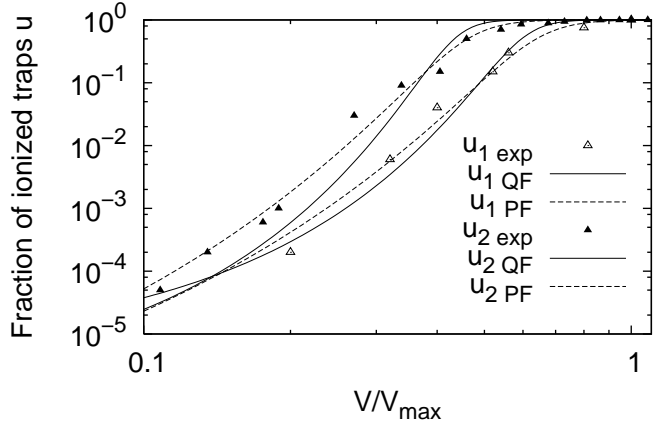}
 \end{center}
 \caption{ \label{fig6}
Fraction of filled traps of the sample in Fig. 5 as function of the applied voltage normalized to the value $V_{max} = 2.5 \ V$ corresponding to the voltage when $u_1=1$ and to the value $V_{max} = 7.4 \ V$ corresponding to the voltage when $u_2=1$ .
Symbols refer to the experimental vaòues deduced using Eq. (8)  from the I-V data in Fig. 5  and lines to the fit using the statistics with the QF and PF models, respectively.
Here  $\Delta \epsilon_{QF} =0.28 \ eV$, $\gamma_{QF}=0.28$,  $\Delta \epsilon_{PF} =0.49 \ eV$ and $\beta_{PF}=4.0 \times 10^{-4} \ (Vm)^{1/2}$ for the case of $u_1$, and $\Delta \epsilon_{QF} =0.32 \ eV$, $\gamma_{QF}=0.11$, and  $\Delta \epsilon_{PF} = 0.51 \ eV$, $\beta_{PF}=2.8 \times 10^{-4} \ (Vm)^{1/2}$ for the case of $u_2$.
}
\end{figure}
\begin{figure}
 \begin{center}
  \includegraphics[width=9cm]{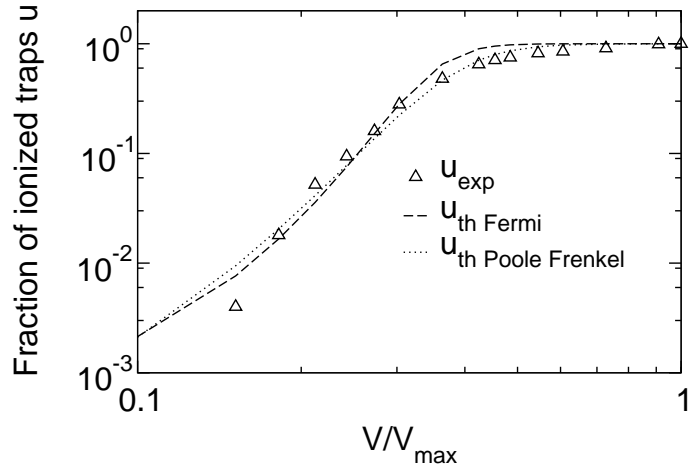}
 \end{center}
 \caption{ \label{fig7}
Fraction of filled traps of the tetracene sample considered in \cite{carbone09} as function of the applied voltage normalized to the value $V_{max} = 3.3 \ V$ corresponding to the voltage when $u=1$ .
Symbols refer to the experimental values deduced using Eq. (8)  from the I-V data of \cite{carbone09}  and lines to the fit using the statistics with the QF and PF models, respectively.
Here  $\Delta \epsilon_{QF} =0.20 \ eV$, $\gamma_{QF}=0.21$,  $\Delta \epsilon_{PF} =0.31 \ eV$ and $\beta_{PF}=2.5 \times 10^{-4} \ (Vm)^{1/2}$. 
}
\end{figure}
\begin{figure}
 \begin{center}
  \includegraphics[width=9cm]{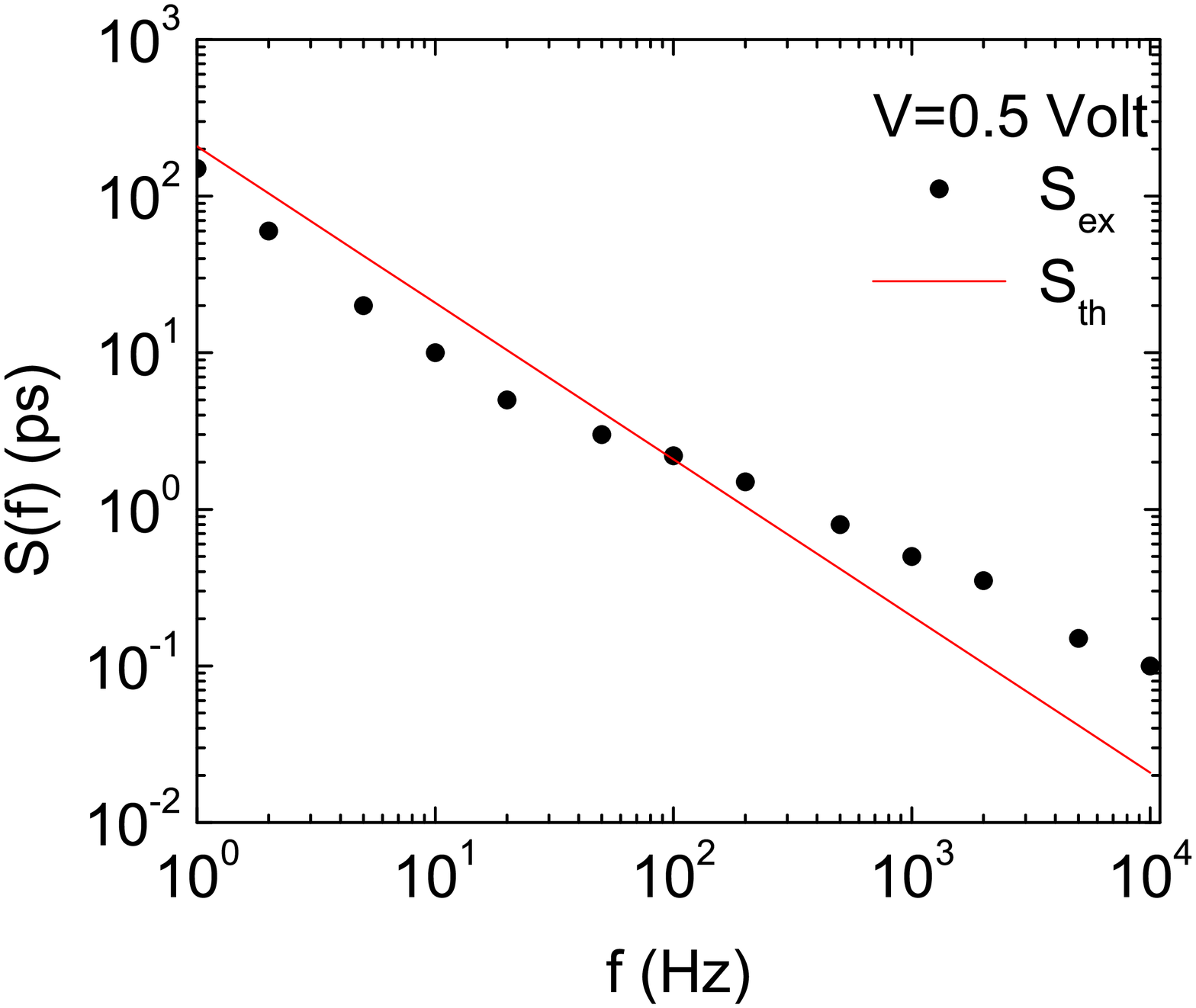}
 \end{center}
 \caption{ \label{fig8}
Relative excess current-noise power spectral density $S(f)$ or the Au/Tc/Al sample of \cite{carbone09} for an applied voltage of 0.5 Volt at room temperature.
Symbols refer to experiments, continuous line to theory (see text).
}
\end{figure}
\begin{figure}
 \begin{center}
  \includegraphics[width=9cm]{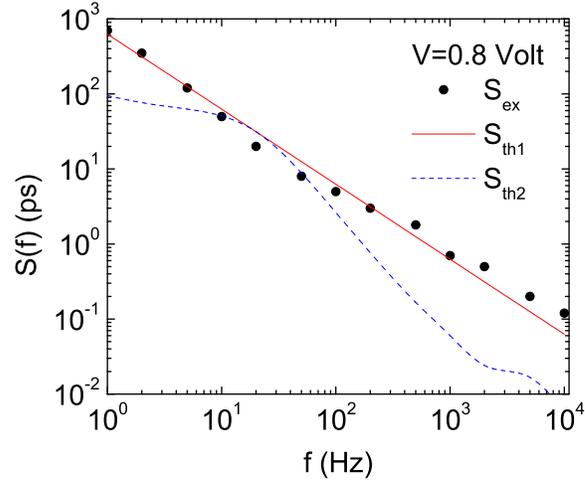}
 \end{center}
 \caption{ \label{fig9}
Relative excess current-noise power spectral density $S(f)$ or the Au/Tc/Al sample of \cite{carbone09} for an applied voltage of 0.8 Volt at room temperature.
Symbols refer to experiments, dashed line refers to theory using a single Lorentzian model for the trapping-detrapping noise, continuous line refers to theory using a broadened distribution of relaxation times  for the trapping-detrapping noise (see text). 
}
\end{figure}
\begin{figure}
 \begin{center}
  \includegraphics[width=9cm]{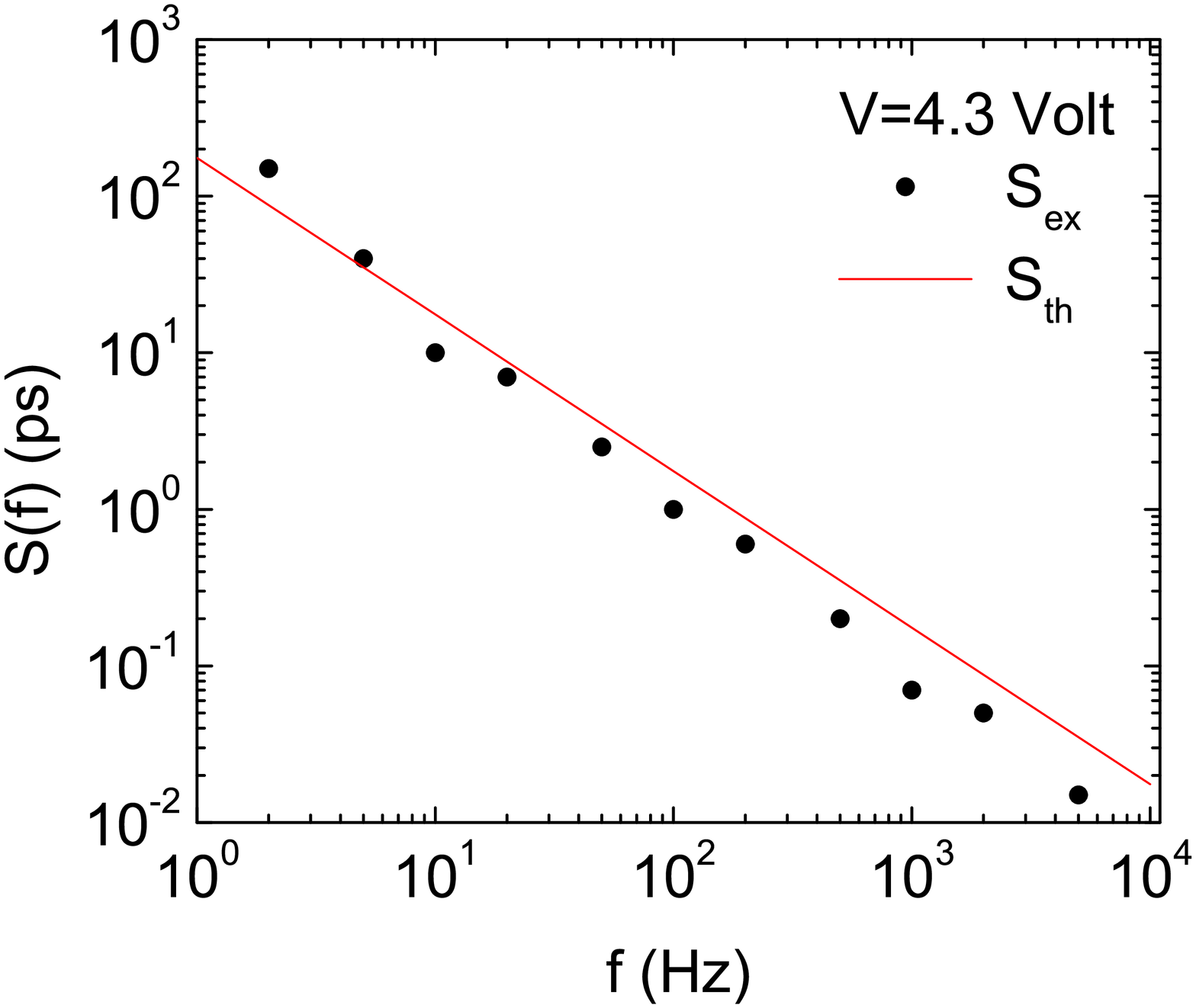}
 \end{center}
 \caption{ \label{fig10}
Relative excess current-noise power spectral density $S(f)$ or the Au/Tc/Al sample of \cite{carbone09} for an applied voltage of 1.5 Volt at room temperature.
Symbols refer to experiments, continuous line to theory.
}
\end{figure}
\begin{figure}
 \begin{center}
  \includegraphics[width=9cm]{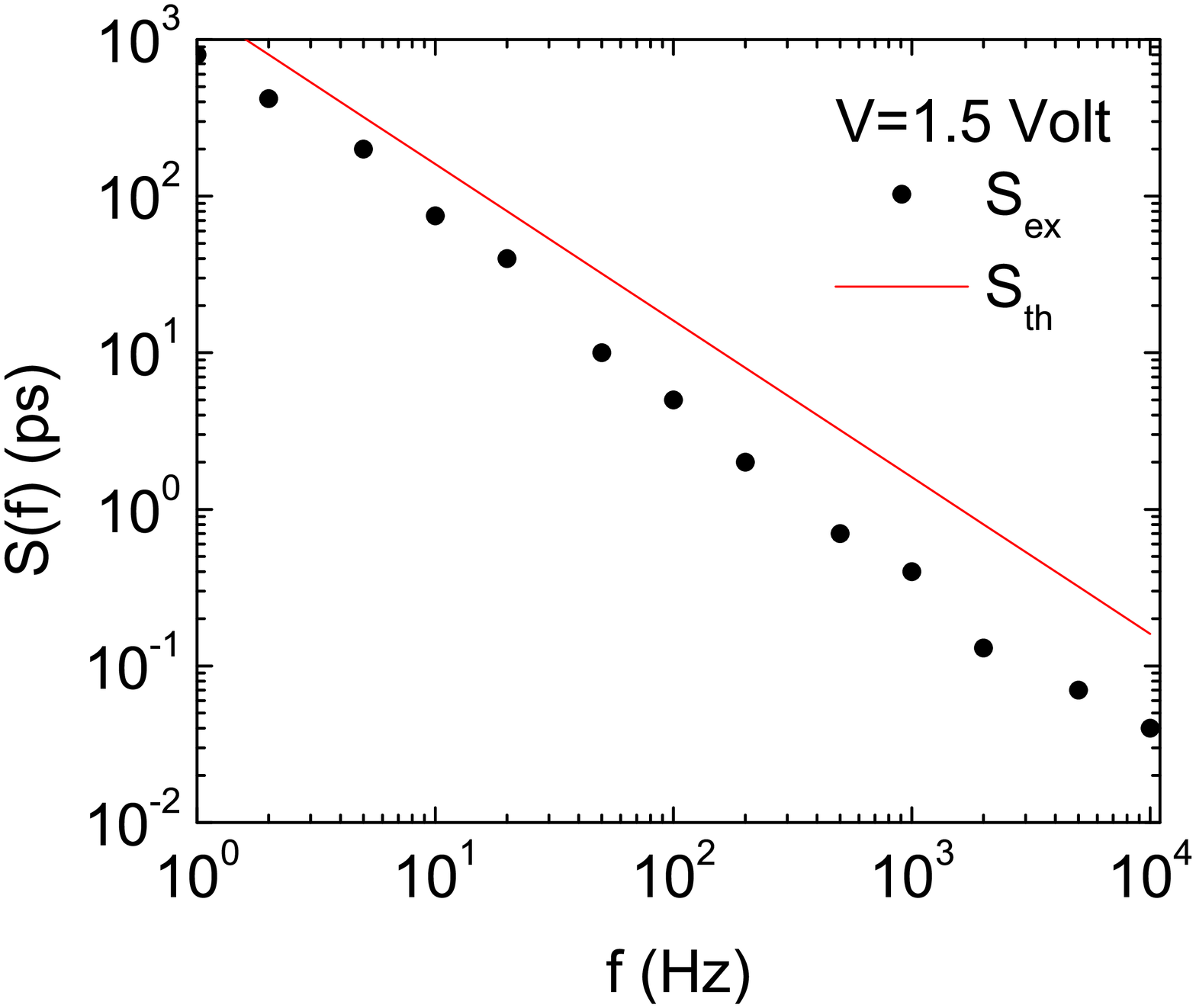}
 \end{center}
 \caption{ \label{fig11}
Relative excess current-noise power spectral density $S(f)$ or the Au/Tc/Al sample of \cite{carbone09} for an applied voltage of 4.3 Volt at room temperature.
Symbols refer to experiments, continuous line to theory.
}
\end{figure}
\begin{figure}
 \begin{center}
  \includegraphics[width=9cm]{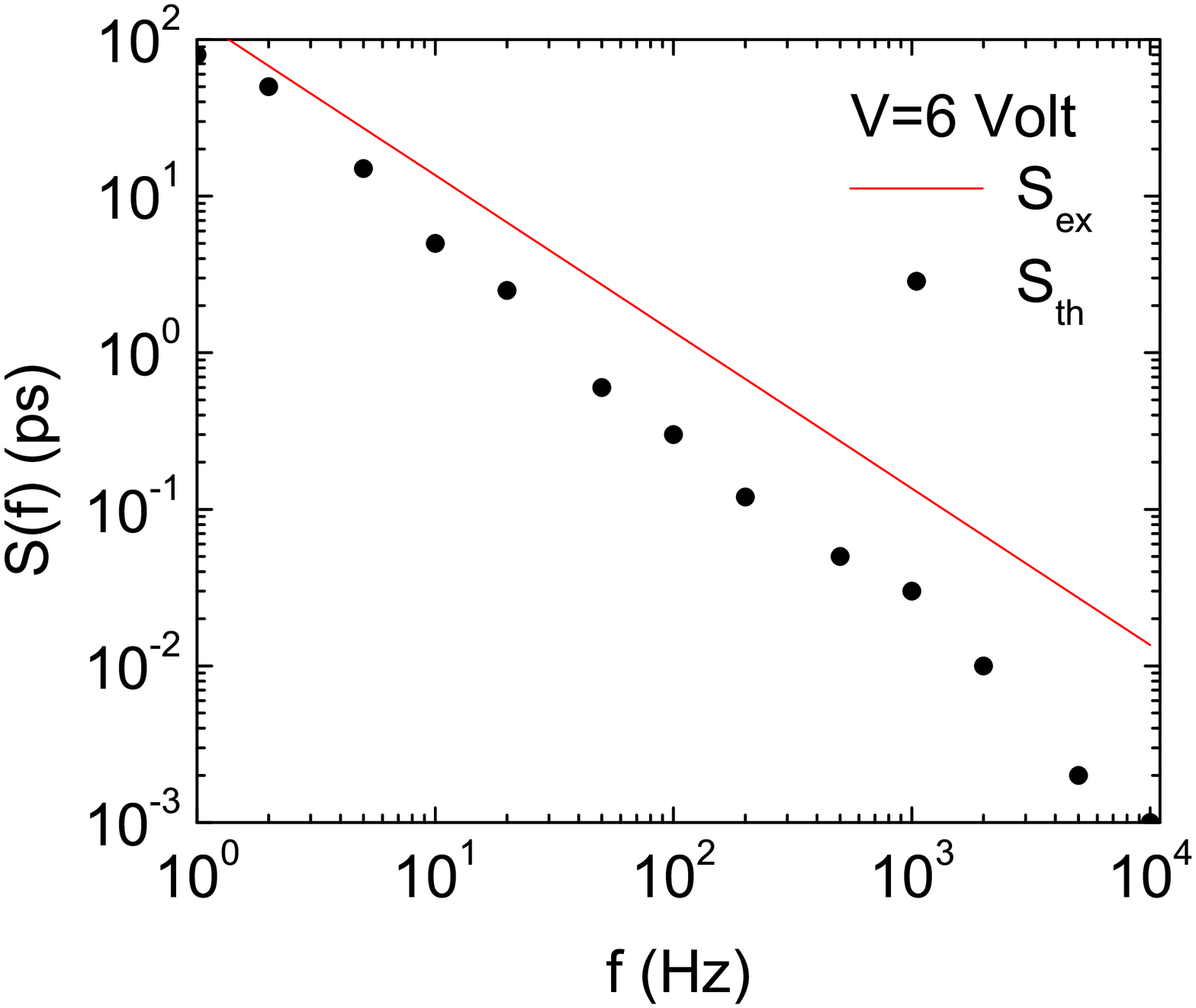}
 \end{center}
 \caption{ \label{fig12}
Relative excess current-noise power spectral density $S(f)$ or the Au/Tc/Al sample of \cite{carbone09} for an applied voltage of 6 Volt at room temperature.
Symbols refer to experiments, continuous line to theory.
}
\end{figure}
\begin{figure}
 \begin{center}
  \includegraphics[width=9cm]{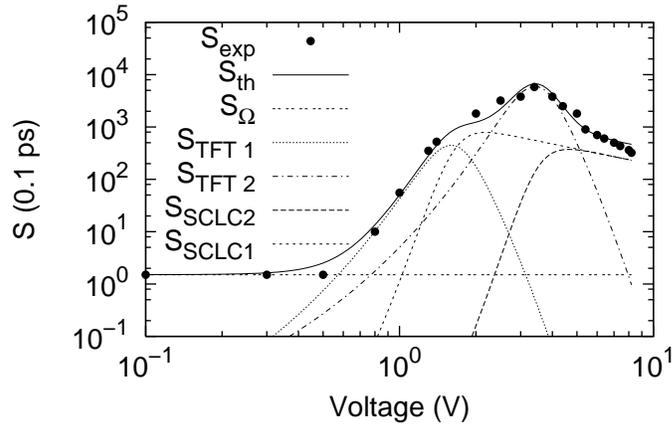}
 \end{center}
 \caption{ \label{fig13}
Relative  excess current-noise at $f=20 \ Hz$ and $T=300 \ K$ vs applied voltage for the pentacene sample of Fig. 5 and Fig. 6.
Symbols refer to experiments.
Dashed lines represent, respectively: (i) the Ohmic noise component at low voltages, (ii) the trapping-detrapping noise component associated with two trapping levels described by fraction of ionized concentrations $u_1$ and $u_2$ at intermediate voltages, (iii) the two Mott-Gurney (SCLC) noise components associated with $u_1=1$ and $u_2=1$ at the highest voltages.
Solid line is obtained by assumming the five noise components according to the decomposition in Eq. (11) and using the fraction of filled traps reported in Fig. 6, see text.
}
\end{figure}
\end{document}